\documentclass[proceedings]{JHEP3}
\usepackage{amsfonts}
\usepackage{amsmath}
\usepackage{epsfig,multicol}

\newbox\mybox

\newcommand\fverb{\setbox\mybox=\hbox\bgroup\verb}
\newcommand\fverbdo{\egroup\medskip\noindent\fbox{\unhbox\mybox}\ }
\newcommand\fverbit{\egroup\item[\fbox{\unhbox\mybox}]}
\conference{Position-dependent noncommutativity}
\abstract{We introduce a new set of noncommutative space-time commutation relations in two space dimensions. The space-space commutation relations are deformations of the standard flat noncommutative space-time relations taken here to have position dependent structure constants. Some of the new variables are non-Hermitian in the most natural choice. We construct their Hermitian counterparts by means of a Dyson map, which also serves to introduce a new metric operator. We
propose $\mathcal{PT}$ like symmetries, i.e.~antilinear involutory
maps, respected by these deformations.
We compute minimal lengths and momenta arising in this space from generalized versions of Heisenberg's uncertainty relations and find that any object in this two dimensional space is string like, 
i.e.~having a fundamental length in one direction
beyond which a resolution is impossible. Subsequently we formulate and partly solve some simple models in these new variables, the free particle, its $\mathcal{PT}$-symmetric deformations and the harmonic oscillator.}

\title{Strings from position-dependent noncommutativity}
\author{Andreas Fring$^\bullet$, Laure Gouba$^\circ$ and Frederik G. Scholtz$%
^{\circ,*}$ \\
$^\bullet$ Centre for Mathematical Science, City University London,\\
$\,\,$ Northampton Square, London EC1V 0HB, UK\\
$^\circ$ National Institute for Theoretical Physics (NITheP), Stellenbosch
7600, South Africa \\
$^*$ Institute for Theoretical Physics, Stellenbosch University, \\
$\,\,$ Stellenbosch 7600, South Africa \\
E-mail: a.fring@city.ac.uk, gouba@sun.ac.za, fgs@sun.ac.za}

\input{tcilatex}

\begin{document}

\section{Introduction}

Noncommutative space-time structures is an old subject dating back over
fifty years to Snyder \cite{Snyder:1946qz}. He introduced noncommutativity
in the hope of regularizing the ultra-violet divergencies that plagued
quantum field theory at that time, but the discovery of renormalization
pushed these ideas on the background. More recently these ideas were revived
by the observation of noncommutativity in certain string theories \cite{wit}
and the compelling arguments for noncommuative space-time structures coming
from gravitational stability \cite{dop}. This has given rise to intense
investigations into noncommutative quantum mechanics, see \cite{delduc} for
an overview, and noncommutative quantum field theories, see \cite%
{Douglas:2001ba,Szabo:2001kg} for reviews.

This noncommutativity is of the simplest possible type, namely, it is
assumed that the Hermitian local coordinates satisfy commutations relations
of the type $[x^{\mu },x^{\nu }]=i\theta ^{\mu \nu }$, with $\theta ^{\mu
\nu }$ a constant antisymmetric tensor. However, there are many other
possibilities that cannot be ruled out by present experimental observation.
Indeed, even in the very first paper by Snyder \cite{Snyder:1946qz}, this
tensor was taken to depend on the coordinates and the momenta. Many
different types of possibilities have been explored since then in so-called $%
\kappa $-Poincar\'{e} noncommutativity \cite{Lukierski:1991pn},
Lie-algebraic approaches \cite{Sasakura:2000vc} , other fuzzy spaces (see 
\cite{bal} for a comprehensive overview) and also more recently in a more
generic position dependent approach \cite{Gomes:2009tk,Gomes:2009rz} that
takes $\theta ^{\mu \nu }$ to be a function of the position coordinates,
i.e.~assuming $\theta ^{\mu \nu }(x)$. In the latter case the consistency of
the Jacobi identity involving one momentum and two coordinate variables also
requires a change in the mutual commutators between positions and momenta
(see e.g. \cite{Gomes:2009tk,Gomes:2009rz}). Relations of such type are
common in a more algebraic approach in which Heisenberg's canonical
commutation relations or the relations between creation and annihilation
operators are directly deformed, e.g.~\cite{Wess1,Brodimas,Kempf2}. As
deformations of this form will almost inevitably lead to non-Hermitian local
coordinates, it was pointed out recently \cite{Bagchi:2009wb} (see also \cite%
{Jana:2009xa}), that therefore these type of structures are related directly
to another subject of current interest, namely non-Hermitian Hamiltonian
systems with real eigenvalues, see e.g.~\cite{Benderrev,Rev3} for recent
reviews on the subject.

The main aim of this manuscript is to explore this interrelation further and
study the consequences of a simple position dependent deformation of the
noncommutative local coordinate commutations relations. Our manuscript is
organized as follows: In section 2 we deform the standard relations of flat
noncommutative space-time and introduce our new version of dynamical
space-time. We argue that the most natural choice for the new variables
leads to non-Hermiticity in one position and one momentum variable. By
constructing a Dyson map we provide their corresponding set of Hermitian
counterparts. We also propose $\mathcal{PT}$-like symmetries respected by
the new set of variables. In section 3 we derive a minimal length and a
minimal momentum resulting from these deformations. In section 4 we study
some simple models formulated in terms of our new set of variables. Our
conclusions are drawn in section 5.

\section{Position dependent noncommutative space-time}

We restrict ourselves here to two dimensional space. The most commonly
investigated noncommutative space-time is flat obeying in this case the
relations%
\begin{equation}
\begin{array}{lll}
\lbrack x_{0},y_{0}]=i\theta , & [x_{0},p_{x_{0}}]=i\hbar ,\qquad \quad & 
[y_{0},p_{y_{0}}]=i\hbar , \\ 
\lbrack p_{x_{0}},p_{y_{0}}]=0,\qquad \quad & [x_{0},p_{y_{0}}]=0, & 
[y_{0},p_{x_{0}}]=0,%
\end{array}
\label{1}
\end{equation}%
with $\theta \in \mathbb{R}$. It is well known that a relation to the
conventional commutative space-time variables can be achieved by the
so-called Bopp-shift in space%
\begin{equation}
x_{0}=x_{s}-\frac{\theta }{\hbar }p_{y_{s}}\qquad \text{and\qquad }%
y_{0}=y_{s},  \label{Bopp}
\end{equation}%
where the standard coordinates $x_{s}$,$y_{s}$ now commute $[x_{s},y_{s}]=0$
and all the remaining commutators remain unchanged when replacing the
subscript $0$ by $s$ in (\ref{1}). Most commonly the Bopp shift is taken to
be more symmetrical as $x_{0}=x_{s}-\theta /(2\hbar )p_{y_{s}}$ and $%
y_{0}=y_{s}+\theta /(2\hbar )p_{x_{s}}$, but for several reasons the
representation (\ref{Bopp}) will be more convenient for our purposes. We
will now explore a simple possibility to deform the relations (\ref{1}) by
introducing a set of new variables $X,Y,P_{x},P_{y}$ of yet unknown
properties and convert the constant $\theta $ into a function $\theta
\rightarrow \theta (X,Y)$, by choosing as one possibility $\theta (X,Y)=$ $%
\theta (1+\tau Y^{2})$. As mentioned above consistency of the Jacobi
identities requires to alter the remaining commutators. We propose here a
simple consistent position dependent and in one case also momentum dependent
deformation of (\ref{1}) satisfying all possible permutations of the Jacobi
identities\footnote{%
That is $\left[ A,[B,C]\right] +\left[ B,[C,A]\right] +\left[ C,[A,B]\right]
=0$ for $A,B,C\in \{X,Y,P_{x},P_{y}\}$.} 
\begin{equation}
\begin{array}{lll}
\lbrack X,Y]=i\theta (1+\tau Y^{2}),\quad \quad & [X,P_{x}]=i\hbar (1+\tau
Y^{2}), & [Y,P_{y}]=i\hbar (1+\tau Y^{2}), \\ 
\lbrack P_{x},P_{y}]=0, & [X,P_{y}]=2i\tau Y(\theta P_{y}+\hbar X),\quad
\quad & [Y,P_{x}]=0.%
\end{array}
\label{3}
\end{equation}%
Obviously by construction we recover the standard flat noncommutative
space-time (\ref{1}) in the limit $\tau \rightarrow 0$. For reasons which
will become apparent below, we also have to take $\tau \geq 0$. We may now
represent the algebra (\ref{3}) in terms of the standard flat Hermitian
noncommutative space-time momentum and position operators $%
x_{0},y_{0},p_{x_{0}},p_{y_{0}}$ as%
\begin{equation}
X=(1+\tau y_{0}^{2})x_{0},\quad Y=y_{0},\quad P_{x}=p_{x_{0}},\quad \text{%
and\quad }P_{y}=(1+\tau y_{0}^{2})p_{y_{0}}.  \label{Xx}
\end{equation}%
From this representation follows immediately that some of the operators
involved are no longer Hermitian. We observe 
\begin{equation}
X^{\dagger }=X+2i\tau \theta Y,\quad Y^{\dagger }=Y,\quad P_{y}^{\dagger
}=P_{y}-2i\tau \hbar Y,\quad \text{and\quad }P_{x}^{\dagger }=P_{x},
\end{equation}%
i.e.~the $X$-coordinate and the momentum in $Y$-direction $P_{y}$ are not
Hermitian. An immediate consequence is that models formulated in terms of
the new set of variables will in general also not be Hermitian. However,
envoking the often synonymously used concepts of quasi-Hermiticity \cite%
{Dieu,Will,Urubu} and pseudo-Hermiticity \cite%
{pseudo1,pseudo2,Mostafazadeh:2002hb} in the context of studying
non-Hermitian systems, one may try to find a similarity transformation,
i.e.~a Dyson map \cite{Dyson}, and convert the non-Hermitian system into a
Hermitian one. Whereas usually this is carried out for some concrete
Hamiltonian, as this is a common starting point, we can, as suggested in 
\cite{Bagchi:2009wb}, perform this here directly for the set of
non-Hermitian observables $\mathcal{O}\neq \mathcal{O}^{\dagger }$, that is
we seek an operator $\eta $ such that%
\begin{equation}
\eta \mathcal{O}\eta ^{-1}=o=o^{\dagger }.  \label{O}
\end{equation}%
For the case at hand we find that the Dyson map can be taken to be $\eta
=(1+\tau Y^{2})^{-1/2}$, such that the new Hermitian variables $%
x,y,p_{x},p_{y}$ in terms of the standard two dimensional flat
noncommutative space-time variables (\ref{1}) become%
\begin{equation}
\begin{array}{ll}
x=\eta X\eta ^{-1}=(1+\tau y_{0}^{2})^{\frac{1}{2}}x_{0}(1+\tau y_{0}^{2})^{%
\frac{1}{2}},~ & p_{x}=\eta P_{x}\eta ^{-1}=p_{x_{0}}, \\ 
y=\eta Y\eta ^{-1}=y_{0}, & p_{y}=\eta P_{y}\eta ^{-1}=(1+\tau y_{0}^{2})^{%
\frac{1}{2}}p_{y_{0}}(1+\tau y_{0}^{2})^{\frac{1}{2}}.%
\end{array}
\label{herm}
\end{equation}%
Our requirement $\tau \geq 0$ ensures here that the Dyson map will be
non-singular. By construction, the algebra satisfied by these variables is
isomorphic to (\ref{3}) 
\begin{equation}
\begin{array}{lll}
\lbrack x,y]=i\theta (1+\tau y^{2}),\quad \quad & [x,p_{x}]=i\hbar (1+\tau
y^{2}), & [y,p_{y}]=i\hbar (1+\tau y^{2}), \\ 
\lbrack p_{x},p_{y}]=0, & [x,p_{y}]=2i\tau y(\theta p_{y}+\hbar x),\quad
\quad & [y,p_{x}]=0.%
\end{array}
\label{laure}
\end{equation}%
Obviously we may also express these variables in terms of the standard
commuting space-time variables $x_{s},y_{s},p_{x_{s}},p_{y_{s}}$ by
utilizing (\ref{Bopp}). in principle we could have written down directly the
Hermitian representation (\ref{herm}) for our deformed algebra and skipped
the introduction of the non-Hermitian variables altogether. However, whereas
the representation (\ref{Xx}) is easy to guess it is not obvious how one
would construct (\ref{herm}) by avoiding the step via the (\ref{Xx}).

As is well established, given $\eta $ we can immediately define a new metric
operator $\rho =\eta ^{\dagger }\eta $ and an inner product $\langle \quad
|\quad \rangle _{\rho }$, which in terms of the standard inner product $%
\langle \quad |\quad \rangle $ is defined as 
\begin{equation}
\langle \Phi |\Psi \rangle _{\rho }:=\langle \Phi |\rho \Psi \rangle ,
\label{inner}
\end{equation}%
for arbitrary states $\langle \Phi |$ and $|\Psi \rangle $. The operators $%
\mathcal{O}$ are then Hermitian with respect to this new metric%
\begin{equation}
\langle \Phi |\mathcal{O}\Psi \rangle _{\rho }=\langle \mathcal{O}\Phi |\Psi
\rangle _{\rho }.
\end{equation}%
Since in our case the Dyson map $\eta $ is Hermitian, the metric operator is
therefore simply computed to be $\rho =\eta ^{2}=(1+\tau Y^{2})^{-1}$.

Alternatively one may also exploit Wigner's observation \cite{EW} about
antilinear operators to investigate the reality of eigenvalue spectra. He
found that operators invariant under such transformations possess real
eigenvalues when in addition their eigenfunctions also respect this
symmetry. $\mathcal{PT}$-symmetry, i.e.~a simultaneous parity transformation 
$\mathcal{P}$ and time reversal $\mathcal{T}$ is a particular examples of
such an operator \cite{Bender:1998ke}. Let us therefore see how $\mathcal{PT}
$-symmetry manifests itself for the above set of variables. We observe for
instance that%
\begin{equation}
\begin{array}{llllll}
\mathcal{PT}\text{: \ \ \ } & x_{s}\mapsto x_{s},~ & y_{s}\mapsto -y_{s},~ & 
p_{x_{s}}\mapsto -p_{x_{s}},~ & p_{y_{s}}\mapsto p_{y_{s}},~ & i\mapsto -i,
\\ 
& x_{0}\mapsto x_{0},~ & y_{0}\mapsto -y_{0},~ & p_{x_{0}}\mapsto
-p_{x_{0}},~ & p_{y_{0}}\mapsto p_{y_{0}},~ & i\mapsto -i, \\ 
& X\mapsto X,~ & Y\mapsto -Y,~ & P_{x}\mapsto -P_{x},~ & P_{y}\mapsto P_{y},~
& i\mapsto -i, \\ 
& x\mapsto x,~ & y\mapsto -y,~ & p_{x}\mapsto -p_{x},~ & p_{y}\mapsto p_{y},~
& i\mapsto -i,%
\end{array}%
\end{equation}%
leaves the commutation relations (\ref{1}), (\ref{3}) and (\ref{laure})
invariant. We have reflected here only in the $y$-direction and left the $x$%
-direction unaltered. If we wish to have a reflection also in the $x$%
-direction, we are forced to change $\theta \mapsto -\theta $. With regard
to the standard $\mathcal{PT}$-transformation this would imply that $\theta $
has to be taken to be purely imaginary, i.e.~$\theta \in i\mathbb{R}$. This
is a quite unappealing variant, as this will imply that we have lost the
Hermiticity of the original flat space variables $x_{0}$ and $y_{0}$. This
option was investigated in \cite{Giri:2008iq}. However, as we argued here
that is not necessary in order to ensure real eigenvalues, which is the
whole purpose of utilizing this symmetry, as this just requires \emph{any}
type of antilinear and involutory operator. In case we would also like to
have a reflection in the $x$-direction, we can alternatively simply define a
new map 
\begin{equation}
\begin{array}{lllllll}
\mathcal{P}_{\theta }\mathcal{T}\text{: \ \ \ } & x_{s}\mapsto -x_{s},~ & 
y_{s}\mapsto -y_{s},~ & p_{x_{s}}\mapsto p_{x_{s}},~ & p_{y_{s}}\mapsto
p_{y_{s}},~ & \theta \mapsto -\theta , & i\mapsto -i, \\ 
& x_{0}\mapsto -x_{0},~ & y_{0}\mapsto -y_{0},~ & p_{x_{0}}\mapsto
p_{x_{0}},~ & p_{y_{0}}\mapsto p_{y_{0}},~ & \theta \mapsto -\theta , & 
i\mapsto -i, \\ 
& X\mapsto -X,~ & Y\mapsto -Y,~ & P_{x}\mapsto P_{x},~ & P_{y}\mapsto P_{y},~
& \theta \mapsto -\theta , & i\mapsto -i, \\ 
& x\mapsto -x,~ & y\mapsto -y,~ & p_{x}\mapsto p_{x},~ & p_{y}\mapsto p_{y},~
& \theta \mapsto -\theta . & i\mapsto -i.%
\end{array}%
\end{equation}%
Clearly the newly defined map $\mathcal{P}_{\theta }\mathcal{T}$ \ is as
required antilinear and involutory, that is $\mathcal{P}_{\theta }\mathcal{T}%
^{2}=\mathbb{I}$. We stress that in this map the minus sign in $\theta
\mapsto -\theta $ is not generated by the antilinear nature of $\mathcal{T}$%
, but is simply imposed on the real $\theta $.

\section{Minimal uncertainties}

As is well known in standard flat space-time noncommutativity (\ref{1}),
Heisenberg's uncertainty principle applied to a simultaneous measurement of $%
x_{0}$ and $y_{0}$, will lead to the fact that they can not be known any
longer both at the same time with absolute precision, but we have to satisfy 
$\Delta x_{0}\Delta y_{0}\geq \theta /2$. However, we can still measure $%
x_{0}$ precisely, that is we can take the limit $\Delta x_{0}\rightarrow 0$,
when we give up any knowledge about the $y_{0}$-direction and allow $\Delta
y_{0}\rightarrow \infty $. The same holds for $x_{0}\leftrightarrow y_{0}$.
The consequences are more severe once the right hand sides of the
commutation relations in (\ref{1}) cease to be constants, but become
functions of the coordinates and/or the momenta. In that case we might
encounter for a particular observable say $A$, that the limit $\Delta
A\rightarrow 0$ can not be carried out without violating the uncertainly
relations, such that $\Delta A$ can never be made smaller than a certain
value $\Delta A_{\min }$ irrespective of what happens to the other variable
involved in the measurement. In that case $A$ can never be known below a
precision of its minimal uncertainty $\Delta A_{\min }$. For the system of
variables satisfying the commutation relations in (\ref{3}) the uncertainty
relations become%
\begin{equation}
\Delta A\Delta B\geq \frac{1}{2}\left\vert \left\langle [A,B]\right\rangle
_{\rho }\right\vert \qquad \ \ \ \ \ \text{for \ }A,B\in \{X,Y,P_{x},P_{y}\},
\label{HU}
\end{equation}%
where we have to employ the inner product as defined in (\ref{inner}).
Starting with a simultaneous $X,Y$-measurement and following the standard
arguments, see e.g.~\cite{Kempf:1993bq,Bagchi:2009wb}, for minimizing the
expression (\ref{HU}) we have to solve 
\begin{equation}
\partial _{\Delta Y}f(\Delta X,\Delta Y)=0\qquad \text{and\qquad\ }f(\Delta
X,\Delta Y)=0,
\end{equation}%
for $\Delta X$ with $f(\Delta X,\Delta Y)$ defined as%
\begin{eqnarray}
f(\Delta X,\Delta Y) &=&\Delta X\Delta Y-\frac{1}{2}\left\vert \left\langle
[X,Y]\right\rangle _{\rho }\right\vert =\Delta X\Delta Y-\frac{\theta }{2}%
\left( 1+\tau \left\langle Y^{2}\right\rangle _{\rho }\right) , \\
&=&\Delta X\Delta Y-\frac{\theta }{2}\left( 1+\tau \left\langle
Y\right\rangle _{\rho }^{2}+\tau \Delta Y^{2}\right) .
\end{eqnarray}%
This leads to a minimal length for $X$ in a simultaneous $X,Y$-measurement 
\begin{equation}
\Delta X_{\min }=\theta \sqrt{\tau }\sqrt{1+\tau \left\langle Y\right\rangle
_{\rho }^{2}}.  \label{xmin}
\end{equation}%
There is no nonvanishing minimal length in $Y$ as we may take the limit $%
\Delta Y\rightarrow 0$ without violating the inequality. This means in the
two dimensional space spanned by $X$ and $Y$ objects are naturally of string
type, being streched out in $X$-direction where a resolution of its
substructure beyond the absolute minimal value $\Delta X_{\min }$, that is $%
\theta \sqrt{\tau }$, is completely impossible. This even holds when
sacrifycing all knowledge about the\ $Y$-direction. On the other hand in
the\ $Y$-direction a complete resolution can be achieved when all
information about the $X$-direction is given up.

Arguing in the same way we do not encounter any minimal length or minimal
momentum in a simultaneous $X,P_{x}$-measurement. However, in a simultaneous 
$Y,P_{y}$-measurement we find a minimal momentum%
\begin{equation}
\Delta \left( P_{y}\right) _{\min }=\hbar \sqrt{\tau }\sqrt{1+\tau
\left\langle Y\right\rangle _{\rho }^{2}},
\end{equation}%
whereas once again there is no minimal length in $Y$.

The argumentation for a simultaneous $X,P_{y}$-measurement is less
straightforward as we encounter terms of the type $\left\langle
YP_{y}\right\rangle _{\rho }$ and $\left\langle YX\right\rangle _{\rho }$,
which can not be treated in the same manner. However, since the behaviour of 
$X$ and $P_{y}$ is linear on both sides of the inequality in both cases, we
do not expect a minimal length or a minimal momentum to arise in this
circumstance.

\section{Models in position dependent noncommutative space-time}

As mentioned, any Hamiltonian depending on the new variables $X$ and $P_{y}$
will obviously no longer be Hermitian. We will now study some examples,
starting by formulating them in terms of these variables, then computing
some equivalent formulations and subsequently solving some of the models in
their most convenient form.

\subsection{The free\ particle}

The simplest Hamiltonian one can envisage in these variable is the free
particle. A priori it is not even clear if the free particle Hamiltonian in
these variables still describes a free particle in the standard sense as the
non-Hermitian nature might alter this property even in this simple case. In
two dimensions the free particle Hamiltonian reads 
\begin{equation}
\mathcal{H}_{\text{f}}(X,Y,P_{x},P_{y})=\frac{1}{2m}(P_{x}^{2}+P_{y}^{2}).
\label{XY}
\end{equation}%
It now depends on our preferences whether we wish to treat the model in
these variables, but with a changed metric as decribed in section 2 or if we
transform the Hamiltonian into standard flat non-commutative space. Using
the relations (\ref{Xx}) we can convert (\ref{XY}) into 
\begin{equation}
\mathcal{H}_{\text{f}}(x_{0},y_{0},p_{x_{0}},p_{y_{0}})=\frac{1}{2m}\left[
p_{x_{0}}^{2}+(1+\tau y_{0}^{2})^{2}p_{y_{0}}^{2}-2i\hbar \tau y_{0}(1+\tau
y_{0}^{2})p_{y_{0}}\right] .  \label{xy}
\end{equation}%
As is apparent this Hamiltonian is non-Hermitian and we still need to change
the metric as decribed above when we wish to compute expectation values or
other physical quantities. Yet another possibility is to map this
Hamiltonian to a Hermitian one, which may then be treated in the
conventional way. In analogy to (\ref{O}) we can achieve this by means of a
similarity transformation. Since all our variables are converted into
Hermitian ones by the same Dyson map, this will also hold for any function
in these variables, as for instance the Hamiltonian. Thus another
possibility to consider $\mathcal{H}_{\text{f}}$ is in terms of the
Hermitian variables introduced in (\ref{herm})%
\begin{equation}
h_{\text{f}}(x,y,p_{x},p_{y})=\eta \mathcal{H}_{\text{f}}\eta ^{-1}=\frac{1}{%
2m}(p_{x}^{2}+p_{y}^{2}).  \label{xy2}
\end{equation}%
By construction this Hamiltonian is Hermitian. Yet another option is to
relate this version to the standard Hermitian flat noncommutative variable
in (\ref{1}). We find 
\begin{eqnarray}
h_{\text{f}}(x_{0},y_{0},p_{x_{0}},p_{y_{0}}) &=&\frac{1}{2m}\left[
p_{x_{0}}^{2}+(1+\tau y_{0}^{2})^{1/2}p_{y_{0}}(1+\tau
y_{0}^{2})p_{y_{0}}(1+\tau y_{0}^{2})^{1/2}\right] ,  \label{xy3} \\
&=&\frac{1}{2m}\left[ p_{x_{0}}^{2}+(1+\tau y_{0}^{2})^{2}p_{y_{0}}^{2}%
\right] +\frac{\hbar ^{2}\tau ^{2}}{m}y_{0}^{2}-\frac{\tau \hbar }{2m}. 
\notag
\end{eqnarray}%
Apparently we have converted the free particle into a harmonic oscillator
like potential in one direction, due to the emergence of the $y_{0}^{2}$%
-term. However, the mixed term in $y_{0}$ and $p_{y_{0}}$ will in fact
compensate for this interaction, having the effect that this Hamiltonian
still allows for a continuous spectrum. According to (\ref{3}) the
non-Hermitian momenta $P_{x},P_{y}$, or likewise $p_{x},p_{y}$, still
commute and therefore we can find simultaneous eigenfunctions for both
operators. Consequently the eigenfunction factorizes $\psi (X,Y)=\varphi
(X)\varphi (Y)$ and since the eigenvalues for $P_{x}^{2}$ are continuous in
the infinite plane, also the spectrum for $P_{y}^{2}$ can not be discrete in
this setting. We should stress that (\ref{XY}), (\ref{xy}), (\ref{xy2}) and (%
\ref{xy3}) are just different points of view to describe the same type of
physics, but care needs to be taken in the selection of meaningful
observables.

Let us now solve this model in its variant (\ref{xy3}). Appealing to the
nonsymmetric Bopp-shift in the form (\ref{Bopp}) and using the fact that in
position space we can represent the momenta as differential operators $%
p_{y_{s}}=-i\hbar \partial _{y_{s}}$, $p_{x_{s}}=-i\hbar \partial _{x_{s}}$
we may re-write the eigenvalue equation 
\begin{equation}
h_{\text{f}}\varphi (x_{0})\varphi (y_{0})=E\varphi (x_{0})\varphi (y_{0})
\end{equation}%
as a decoupled second order differential equation in the two variables $%
x_{0} $ and $y_{0}$%
\begin{equation}
-\frac{\hbar ^{2}}{2m}\left[ \partial _{x_{0}}^{2}+(1+\tau
y_{0}^{2})^{2}\partial _{y_{0}}^{2}-2\tau ^{2}y_{0}^{2}+\frac{\tau }{\hbar }%
\right] \varphi (x_{0})\varphi (y_{0})=E\varphi (x_{0})\varphi (y_{0}).
\label{diff}
\end{equation}%
Here lies another reason for adopting the non-symmetric form (\ref{Bopp}).
This form still guarantees the decoupling of the two sets of variables,
whereas the more symmetric version will lead to a mixing of the $x_{0}$ and $%
y_{0}$ variables. Equation (\ref{diff}) is solved by%
\begin{eqnarray}
\varphi (x_{0}) &=&c_{1}\sin kx_{0}+c_{2}\cos kx_{0}  \label{fx} \\
\varphi (y_{0}) &=&\sqrt{1+\tau y_{0}^{2}}\left[ \tilde{c}_{1}P_{1}^{\mu
}\left( iy_{0}\sqrt{\tau }\right) +\tilde{c}_{2}Q_{1}^{\mu }\left( iy_{0}%
\sqrt{\tau }\right) \right]
\end{eqnarray}%
with continuous eigenenergy%
\begin{equation}
E(k)=\frac{k^{2}\hbar ^{2}}{2m}+\frac{\tau \hbar }{2m},  \label{E}
\end{equation}%
parameterized by $k\in \mathbb{R}$ and $\mu =\sqrt{3+k^{2}/\tau }$. The
functions\ $P_{\nu }^{\mu }\left( x\right) $ and $Q_{\nu }^{\mu }\left(
x\right) $ are associated Legendre polynomials of the first and second kind,
respectively, and $c_{1},c_{2},\tilde{c}_{1},\tilde{c}_{2}$ are integration
constants. Notice that the limit $\tau \rightarrow 0$ to the undeformed case
is nontrivial in this case as we had to introduce a variable transformation
involving $1/\sqrt{\tau }$ in order to convert (\ref{diff}) into the
conventional form of the differential equation solvable by associated
Legendre polynomials.

\subsection{$\mathcal{PT}$ and $\mathcal{P}_{\protect\theta }\mathcal{T}$%
-extensions of the free\ particle}

As explained above we still have a good chance to have well defined models
with real eigenvalues when our Hamiltonian remains invariant with respect to
an antilinear involutory symmetry. Let us therefore in the spirit of
deforming Hermitian models add some additional terms to the free particle
Hamiltonian 
\begin{equation}
\mathcal{H}_{\text{fPT}}(X,Y,P_{x},P_{y})=\frac{1}{2m}(P_{x}^{2}+P_{y}^{2})+%
\lambda (iY)^{n}P_{y}^{m}\text{,\qquad with }n,m\in \mathbb{N}_{0},\lambda
\in \mathbb{R}.  \label{fpt}
\end{equation}%
Clearly, this Hamiltonian remains invariant with respect to the $\mathcal{PT}
$ as well as the $\mathcal{P}_{\theta }\mathcal{T}$-symmetry, i.e.~$[%
\mathcal{PT},\mathcal{H}_{\text{fPT}}]=[\mathcal{P}_{\theta }\mathcal{T},%
\mathcal{H}_{\text{fPT}}]=0$. Since $P_{x}$ also commutes with the added
term, we may still apply the argument of the previous subsection and
construct simultaneous eigenstates for $P_{x}^{2}$ and the remaining term.
For instance, for $n=m=1$ the corresponding differential equation in
position space becomes \ 
\begin{equation}
-\frac{\hbar ^{2}}{2m}\left[ \partial _{x_{0}}^{2}+(1+\tau
y_{0}^{2})^{2}\partial _{y_{0}}^{2}-2\tau ^{2}y_{0}^{2}+\lambda y_{0}(1+\tau
y_{0}^{2})\partial _{y_{0}}\right] \varphi (x_{0})\varphi (y_{0})=E\varphi
(x_{0})\varphi (y_{0}),
\end{equation}%
where we take $E$ directly in the form (\ref{E}). The solution for $\varphi
(x_{0})$ will remain the same, whereas $\varphi (y_{0})$ results now to%
\begin{equation}
\varphi (y_{0})=\left( 1+\tau y_{0}^{2}\right) ^{\kappa }\left[ \tilde{c}%
_{1}P_{\nu }^{\mu }\left( iy_{0}\sqrt{\tau }\right) +\tilde{c}_{2}Q_{\nu
}^{\mu }\left( iy_{0}\sqrt{\tau }\right) \right]
\end{equation}%
with 
\begin{equation}
\nu =\frac{1}{2}+\frac{\lambda }{4\tau },\quad \mu =\frac{\sqrt{\lambda
^{2}+2\lambda \tau +9\tau ^{2}}}{2\tau }-\frac{1}{2}\quad \text{and\quad }%
\kappa =\frac{\sqrt{\lambda ^{2}+4\lambda \tau +4\tau (k^{2}+3\tau )}}{2\tau 
}.  \label{abc}
\end{equation}%
We can take here two different points of view: On one hand we may choose $%
\lambda $ to be a generic constant, which would imply that the model (\ref%
{fpt}) still remains non-Hermitian in the limit $\tau \rightarrow 0$. On the
other hand we can identify $\lambda =\tau $, such that the limit $\tau
\rightarrow 0$ will reduce $\mathcal{H}_{\text{fPT}}$ to a Hermitian
Hamiltonian. In that case the constants in (\ref{abc}) simplify to $\nu =3/4$%
, $\mu =\sqrt{3}-1/2$ and $\kappa =1/2\sqrt{17+4k^{2}/\tau }$. Once again
the limit $\tau \rightarrow 0$ is nontrivial.

\subsection{The harmonic oscillator}

The next natural complication of our previous examples would be the two
dimensional harmonic oscillator%
\begin{equation}
\mathcal{H}_{\text{ho}}(X,Y,P_{x},P_{y})=\frac{1}{2m}(P_{x}^{2}+P_{y}^{2})+%
\frac{m\omega ^{2}}{2}(X^{2}+Y^{2}).
\end{equation}%
Obviously also this model can be re-written in terms of the flat commuting
variables obeying (\ref{1})%
\begin{eqnarray}
\mathcal{H}_{\text{ho}}(x_{0},y_{0},p_{x_{0}},p_{y_{0}}) &=&\frac{1}{2m}%
\left[ p_{x_{0}}^{2}+(1+\tau y_{0}^{2})^{2}p_{y_{0}}^{2}-2i\hbar \tau
y_{0}(1+\tau y_{0}^{2})p_{y_{0}}\right] \\
&&+\frac{m\omega ^{2}}{2}\left[ (1+\tau y_{0}^{2})^{2}x_{0}^{2}+2i\theta
\tau y_{0}(1+\tau y_{0}^{2})x_{0}+y_{0}^{2}\right] .  \notag
\end{eqnarray}%
Since $\mathcal{H}_{\text{ho}}$ is evidently non-Hermitian, we have to
employ a similarity transformation and convert it in the same spirit as in
the previous section into a Hermitian Hamiltonian%
\begin{equation}
h_{\text{ho}}(x,y,p_{x},p_{y})=\frac{1}{2m}(p_{x}^{2}+p_{y}^{2})+\frac{%
m\omega ^{2}}{2}(x^{2}+y^{2}).
\end{equation}%
Using the representation (\ref{herm}) we may of course also re-express this
Hamiltonian in term of the flat commuting variables obeying (\ref{1}) 
\begin{eqnarray}
h_{\text{ho}}(x_{0},y_{0},p_{x_{0}},p_{y_{0}}) &=&\frac{1}{2m}\left[
p_{x_{0}}^{2}+(1+\tau y_{0}^{2})^{1/2}p_{y_{0}}(1+\tau
y_{0}^{2})p_{y_{0}}(1+\tau y_{0}^{2})^{1/2}\right]  \notag \\
&&+\frac{m\omega ^{2}}{2}\left[ (1+\tau y_{0}^{2})^{1/2}x_{0}(1+\tau
y_{0}^{2})x_{0}(1+\tau y_{0}^{2})^{1/2}+y_{0}^{2}\right] ,  \notag \\
&=&\frac{1}{2m}\left[ p_{x_{0}}^{2}+(1+\tau y_{0}^{2})^{2}p_{y_{0}}^{2}%
\right] +\frac{m\omega ^{2}}{2}\left[ 1-2\theta ^{2}\tau ^{2}-2\frac{\hbar
^{2}\tau ^{2}}{m^{2}\omega ^{2}}\right] y_{0}^{2}~~~~ \\
&&+\frac{m\omega ^{2}}{2}(1+\tau y_{0}^{2})^{2}x_{0}^{2}+2i\tau y_{0}(1+\tau
y_{0}^{2})\left[ m\omega ^{2}\theta x_{0}-\frac{\hbar }{m}p_{y_{0}}\right] 
\notag \\
&&-\frac{\tau }{2}\left( m\omega ^{2}\theta ^{2}+\frac{\hbar }{m}\right) . 
\notag
\end{eqnarray}%
Clearly this is a far more complicated model to solve with the same method
as in the previous sections, as the system viewed as a differential equation
no longer decouples in $x_{0}$ and $y_{0}$. We leave the construction of
solutions for this model by alternative means for future work.

\section{Conclusions}

We have provided a new version of noncommutative space-time in two
dimensions, which is dynamical in the sense that the $x,y$-commutation
relations acquire a position dependent structure constant. An immediate
consequence of this deformation of the common flat commutation relations was
that some of the natural variables associated to these new commutation
relations are non-Hermitian. As we have shown this is not dictated by the
commutation relations themselves as there exist an isomorphic algebra in
terms of Hermitian operators (\ref{laure}). However, these variables do not
constitute the natural starting point and only emerge when the Dyson map has
been constructed. This in turn will also give rise to a new natural metric,
which has to be used to compute physical quantities. We encounter here the
well known problem that this metric might not be unique and there could be
other possibilities related to different types of physical observables.

Previous attempts \cite{Gomes:2009tk,Gomes:2009rz} to construct dynamical
deformations of (\ref{1}) were based on a quantization procedure of a
concrete classical system in the presence of constraints, thus providing a
nice physical scenario in which such type of deformed spaces might arise.
However, the resulting algebra is only valid in tems of Dirac brackets up to
the imposed constraints, whereas our algebra (\ref{1}) is selfconsistent,
model independent and entirely placed into a quantum mechanical setting.

The interesting physical consequence we found is that any object in this two
dimensional space will be string like, as we found that one direction is
inevitably bounded by the quantity $\theta \sqrt{\tau }$, beyond which any
further localization is not only impossible but even meaningless. As the
constant value indicates, i.e.~being explicitly dependent on $\tau $, this
is a direct consequence of our very starting point, namely the position
dependent deformation.

We have also analysed the various possibilities to implement $\mathcal{PT}$%
-symmetry and generalized versions of it for our deformed noncommutative
space-time. Our proposed maps have even consequences for the flat version of
it, as we argue that one does not have to take $\theta $ to be purely
imaginary. This avoids the unappealing feature that in doing this one looses
the Hermiticity of $x_{0}$ and $y_{0}$. Having the two types of antilinear
involutory maps, $\mathcal{PT}$ and $\mathcal{P}_{\theta }\mathcal{T}$,
would allow us also to investigate further extensions of any model. Assuming
for some function $f(y,p_{x},p_{y})$ that $[\mathcal{PT},f(y,p_{x},p_{y})]=0$
or $[\mathcal{P}_{\theta }\mathcal{T},f(y,p_{x},p_{y})]=0$ we have now the
option to add terms of the type $x^{n}f(y,p_{x},p_{y})$ or $%
(ix)^{n}f(y,p_{x},p_{y})$, respectively, without violating this symmetry.

Clearly there are many interesting immediate problems arising from our
investigations, such as the investigation of further possibilities of
consistent deformations, the constuction of the solution for the harmonic
oscillator, the study of additional models in terms of our newly proposed
variables, deformations of dispersion relations resulting from the models
considered here and generalizations to fully fledged field theory setting.

\medskip

\noindent \textbf{Acknowledgments:} A.F. would like to thank the National
Institute for Theoretical Physics of South Africa and the Stellenbosch
Institute of Advanced Study for kind hospitality and financial support. L.G.
and F.G.S. are supported under the grant of the National Research Foundation
of South Africa.


\end{document}